# Demonstration of Returning Thouless Pump in a Berry Dipole System

Qingyang Mo[1]*, Shanjun Liang[2]*, Xiangke Lan[2], Jie Zhu[3‡], Shuang Zhang[1,4,5,6,7†]

[1] *New Cornerstone Science Laboratory, Department of Physics, University of Hong Kong; 999077, Hong Kong, China*

[2] *School of Professional Education and Executive Development, The Hong Kong Polytechnic University, Hong Kong, China*

[3] *Institute of Acoustics, School of Physics Science and Engineering, Tongji University, 200092 Shanghai, China*

[4] *State Key Laboratory of Optical Quantum Materials, University of Hong Kong; 999077, Hong Kong, China*

[5] *Department of Electrical & Electronic Engineering, University of Hong Kong; 999077, Hong Kong, China*

[6] *Quantum Science Center of Guangdong-Hong Kong-Macao Great Bay Area, 3 Binlang Road, Shenzhen, China*

[7] *Materials Innovation Institute for Life Sciences and Energy (MILES), HKU-SIRI, Shenzhen, P.R. China*

The Thouless pump, a cornerstone of topological physics, enables unidirectional quantized wave/particle transport via geometric Berry phase engineering in periodically driven systems. While decades of research have been dedicated to monopole-mediated pumping, mechanisms governed by higher-order singularities like Berry dipoles remain unexplored. Here, we report the experimental demonstration of a Berry-dipole-mediated returning Thouless pump (RTP) in a 1D acoustic waveguide array achieved through adiabatic encircling of a Berry dipole singularity. During this adiabatic cycle, an initial edge-localized mode first delocalizes into the bulk and eventually returns to the original edge, marking the characteristic signature of the RTP. Notably, this RTP exhibits an interesting feature of pseudospin flipping. The demonstrated RTP contrasts sharply with the well-studied monopole-governed pumps that feature unidirectional transport.

Thouless pumps represent a fundamental quantum pumping mechanism that connects quantized charge transport in one-dimensional (1D) dynamical systems with higher-dimensional topological phenomena [1–6]. The Rice-Mele model stands out as a prominent platform for studying Thouless pumps, where the pumping mechanism relies on the adiabatic modulation of alternating on-site potentials ($\Delta$) and intra/inter-coupling differences ($\delta$). Combining these two parameters with a momentum dimension, a 3D synthetic space is created housing a Berry monopole—a 3D topological singularity that emits quantized Berry flux [Fig. 1(a)]. Along a closed adiabatic path encircling the Berry monopole in the $\Delta-\delta$ parameter space, the accumulated quantized Zak phase, equivalent to the Chern number obtained through the integral of Berry curvature over the red torus in Fig. 1(a), governs the precise integer-quantized electron transfer between edges during a pumping cycle [refer to Fig. 1(b)]. This phenomenon elucidates how a low-dimensional topological pump corresponds to a high-dimensional topological singularity, a concept validated across diverse platforms such as ultracold atoms [7–10] and photonic/acoustic waveguides [11–18]. Moreover, the realm of Thouless pumps has expanded to encompass higher dimensions [19–21], non-Abelian regimes [22–27], and nonlinear systems [28–35], solidifying its status as a universal framework for topological physics.

Beyond conventional symmetry classifications, emerging higher-order singularities like Berry dipoles are reshaping topological band theory [36–39]. The Berry dipole hosts a dipolar Berry curvature field distribution and usually emerges in quadratic band degeneracies [40] or linear multiband nodes [41]. These singularities herald a new paradigm of delicate topology, characterized by the Hopf and the returning Thouless pump invariants. Furthermore, the dipolar Berry flux enables phenomena inaccessible to monopole-dominated systems, such as 3D Hopf insulators [42–48], oriented Landau levels [40,41,49], chiral-symmetric surface states [50], and 2D returning pumps with spin reversal [51]. These advances raise a pivotal question: *Can Berry dipoles host 1D topological pumps, and how do their quantization rules and dynamics differ from monopole-governed systems?*

Here, we report the experimental demonstration of a Berry-dipole-mediated topological pump, referred to as "returning Thouless pump" (RTP), in a 1D acoustic waveguide array. We construct a Berry dipole singularity in a 3D synthetic space through engineering the couplings between waveguides [Fig. 1(c)]. Along a closed adiabatic path encircling the Berry dipole, RTP transfers the quantized electron from edge to bulk in the red sub-cycle and then returns it to the original edge with spin inversion in the blue sub-cycle [Fig. 1(d)]. These unconventional pumping dynamics are directly observed through the adiabatic evolution of acoustic modes in our waveguide array, i.e., from an edge-localized pseudospin-down ($|\downarrow\rangle$) mode to bulk delocalization and finally a pseudospin-up ($|\uparrow\rangle$) mode localized at the original edge. This pseudospin flipping arises from a locking between Berry curvature field and pseudospin. RTP stands in stark contrast to conventional Thouless pumps governed by Berry monopoles, where wave/particle undergoes unidirectional quantized transport.

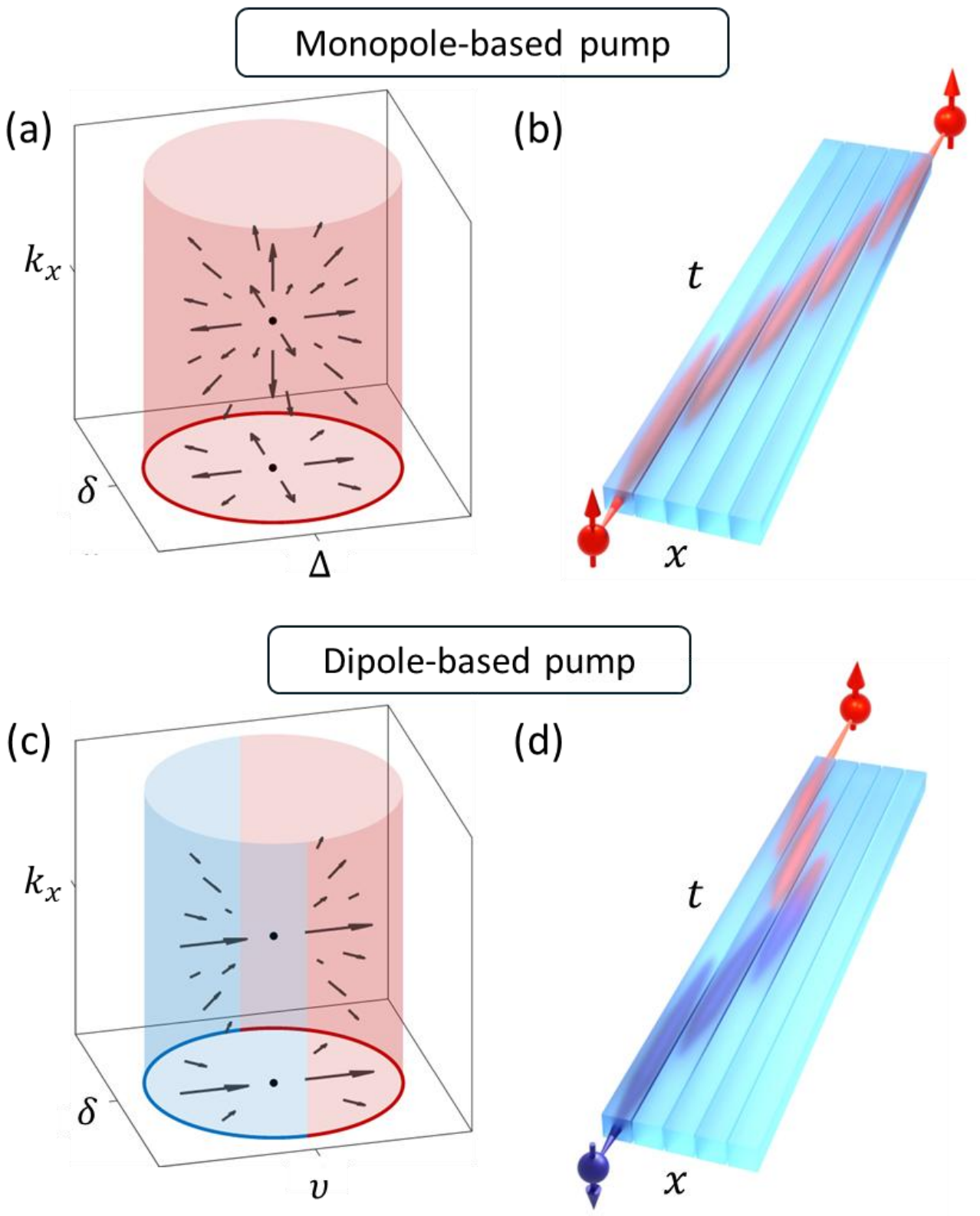

FIG. 1. Comparison between monopole- and dipole-governed pumps. (a) The Berry monopole in the synthetic space. (b) The Thouless pump along the red circle in (a). (c) The Berry dipole in the synthetic space. (d) The returning Thouless pump (RTP) along the circle in (c).

*Tight-binding model*. We start by proposing a 1D lattice model for RTP as shown in Fig. 2(a), whose Hamiltonian can be written as

$$\hat{H}=\sum_j\{\omega_1\hat{b}_j^\dagger\hat{c}_j+\omega_2\hat{c}_j^\dagger\hat{b}_{j+1}+\upsilon\hat{a}_j^\dagger\hat{b}_j+h.c.\}, \tag{1}$$

where $\hat{a}_j^\dagger(\hat{a}_j)$, $\hat{b}_j^\dagger(\hat{b}_j)$, and $\hat{c}_j^\dagger(\hat{c}_j)$ are the creation (annihilation) operators acting on the "A", "B", and "C" sites of the $j$-th unit cell. The intra- and inter-cell hoppings between B and C sites are parameterized as $\omega_1=\omega_0-\delta/2$ and $\omega_2=\omega_0+\delta/2$, respectively, while $\upsilon$ governs the A-B coupling. Combined with two parameter dimensions $(\delta,\upsilon)$ and one momentum dimension ($k_x$), this model synthesizes a 3D parameter space hosting a triply degenerate nodal point at $(k_x,\delta,\upsilon)=(\pi/2,0,0)$. The Berry curvature field near this singularity adopts a standard dipolar configuration as

$$\Omega=\frac{(\vec{d}\cdot\vec{q})\vec{q}}{|\vec{q}|^4}, \tag{2}$$

where $\vec{q}=(\omega_0(k_x-\pi/2),\delta,\upsilon)$ defines a synthetic momentum vector, and $\vec{d}=(0,0,1)$ is the dipole moment vector, as shown in Fig. 2(b) (see more details of the Berry dipole in Section 1 of [52]). The effective Hamiltonian for this Berry dipole preserves chiral symmetry and a rotation symmetry about the $\upsilon$-axis [41], resulting in vanishing integral of Berry curvature over a closed torus enclosing a Berry dipole ($\oint\Omega\cdot dS=0$). Nonetheless, the intrinsic quantization of the dipole singularity, referred as "RTP invariant" [37,51], can be revealed by partitioning the surface into red ($\upsilon>0$) and blue ($\upsilon<0$) sub-tori in Fig. 1(c):

$$Q=\frac{1}{2\pi}\int_{\upsilon\gtrless0}\Omega\cdot dS=\begin{cases}\pm\frac{1}{2} & \text{for 1st and 3rd bands}\\ \mp1 & \text{for the 2nd band}\end{cases}, \tag{3}$$

indicating that the dipole comprises a pair of coalesced Berry monopoles with opposite chirality.

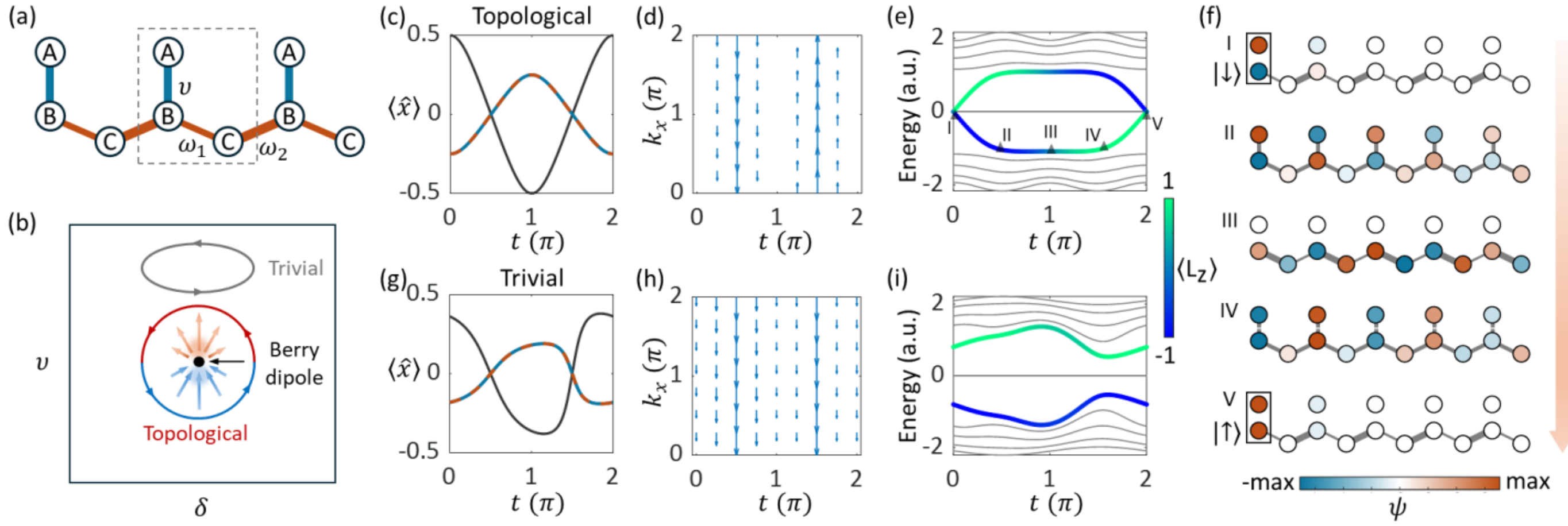

FIG. 2. (a) The 1D lattice model for RTP. (b) The dipolar Berry curvature field in the $\delta-\upsilon$ plane with $k_x=\pi$. The loops enclosing or not enclosing the Berry dipole denote nontrivial and trivial evolution processes, respectively. (c-e) The motion of Wannier centers (WCs) (c), pseudospin distribution for the 1st/2nd /3rd (red/black/blue line) bands (d), and energy spectrum

(e) during the topological pumping process (red loop in(b)), where $\omega_0 = 1$, $\delta = cos(t)$, and $\upsilon = sin(t)$. (f) The field distributions for the lower edge band with $t = 0, \pi/2, \pi, 3\pi/2, 2\pi$ (i–v). The thickness of connecting tubes represent the strength of couplings. Solid and dashed tubes denote positive and negative couplings, respectively. (g-i) The motion of WCs (g), pseudospin distribution for the 1st/2nd /3rd (red/black/blue line) bands (h), and energy spectrum (i) during the trivial evolution (gray loop in (b)), where $\omega_0 = 1$, $\delta = cos(t)$, and $\upsilon = 0.8 + 0.3\, sin(t)$. The color of the edge states in (e, i) represent the pseudospin polarization.

This quantized RTP invariant can be directly linked to the motion of the Wannier centers (WCs) in our RTP model. We consider an adiabatic pumping protocol along a closed elliptical trajectory in the $\delta - \upsilon$ parameter plane [red/blue loop in Fig. 2(b)], mathematically defined as

$$\delta(t) = \delta_0 \cdot \cos(t), \upsilon(t) = \upsilon_0 \cdot \sin(t), t \in (0, 2\pi), \quad (4)$$

which encloses the Berry dipole singularity. As shown in Fig. 2(c), the WCs exhibit bidirectional motion, traversing 1/2 (1st and 3rd bands) or 1 (2nd band) unit cell twice per full pumping cycle. They first propagate rightward (leftward) by a quantized 1/2 (1) unit cell during $\upsilon > 0$ $(t \in [0, \pi])$, then return to their initial positions during $\upsilon < 0$ $(t \in [\pi, 2\pi])$, corresponding to opposite Berry fluxes in Eq. (3). Furthermore, due to spin-momentum locking, the pseudospin polarization ($\langle L_z \rangle$) is reversed when crossing from $\upsilon > 0$ to $\upsilon < 0$ subspaces, as shown in Fig. 2(d) (see more details in Section 2 of [52]). The opposite motion direction and pseudospin $\upsilon \gtrless 0$ subspaces as shown in Fig. 1(d) lead to the unconventional RTP.

This RTP dynamics can be explicitly revealed in a finite 5-cell lattice. As shown in Fig. 2(e), the energy spectrum displays chiral-symmetric edge states (red curves) that linearly intersect at $t = 0$. Fig. 2(f) demonstrates a complete pseudospin-inversion cycle. Initially ( $t = 0$ ), an antisymmetric edge mode ($|\downarrow\rangle$ mode) localizes at the left edge, characterized by opposite field distributions on the A/B sublattices [Fig. 2(f-i)]. As the pumping progresses, this mode gradually delocalizes into bulk [Fig. 2(f-ii)–(iv)] and flips the pseudospin during $\pi/2 < t < 3\pi/2$ , before reconverging at the original edge by $t = 2\pi$ as a symmetric mode ($|\uparrow\rangle$ mode) with uniform sublattice occupation [Fig. 2(f-v)]. Noting that the RTP remains robust against continuous deformation of evolutionary trajectory as long as it encircles the Berry dipole (see more details in Section 3 of [52]).

The essential role of dipole topology in RTP is confirmed by contrasting trajectories that exclude the singularity. For parameter paths encircling trivial regions [gray loop in Fig. 2(b)], the WCs oscillate locally without traversing the unit cell, as shown in Fig. 2(g). Meanwhile, the pseudospin polarization for the lowest band remain the same during the whole process [Fig. 2(h)]. The edge states remain gapped throughout the cycle [Fig. 2(i)], which forces the system to return to its initial state with preserved pseudospin. The distinct evolution results in nontrivial/trivial circles conclusively demonstrate that the RTP dynamics are intrinsically tied to the dipole's topological structure.

*Acoustic waveguide design*. Guided by the above tight-binding model, we design an array of coupled acoustic waveguides with carefully tailored coupling configurations to realize RTP. Fig. 3(a) illustrates the cross section of unit cell structure, where three rectangular waveguides (labeled as A, B, and C) correspond to the three sublattices in the tight-binding model (see more details in Section 4 of [52]). The narrow tube connecting A and B waveguides corresponds to the coupling $\upsilon$, while intra/inter-cell tubes connecting B and C waveguides denote the couplings $\omega_1/\omega_2$, respectively. Only the $p$ mode in the waveguides is considered, hosting an antisymmetric field distribution along the long side of each waveguide. Hence, both sign and amplitude of coupling coefficients can be controlled by the displacement between the centers of waveguides and the connecting tubes. With a fixed configuration of connecting tubes, the misalignments of centers of A and C waveguides $s_{1,2}$ correspond to the pumping parameter $\upsilon$ and $\delta$, respectively (see more details in Section 5 of [52]).

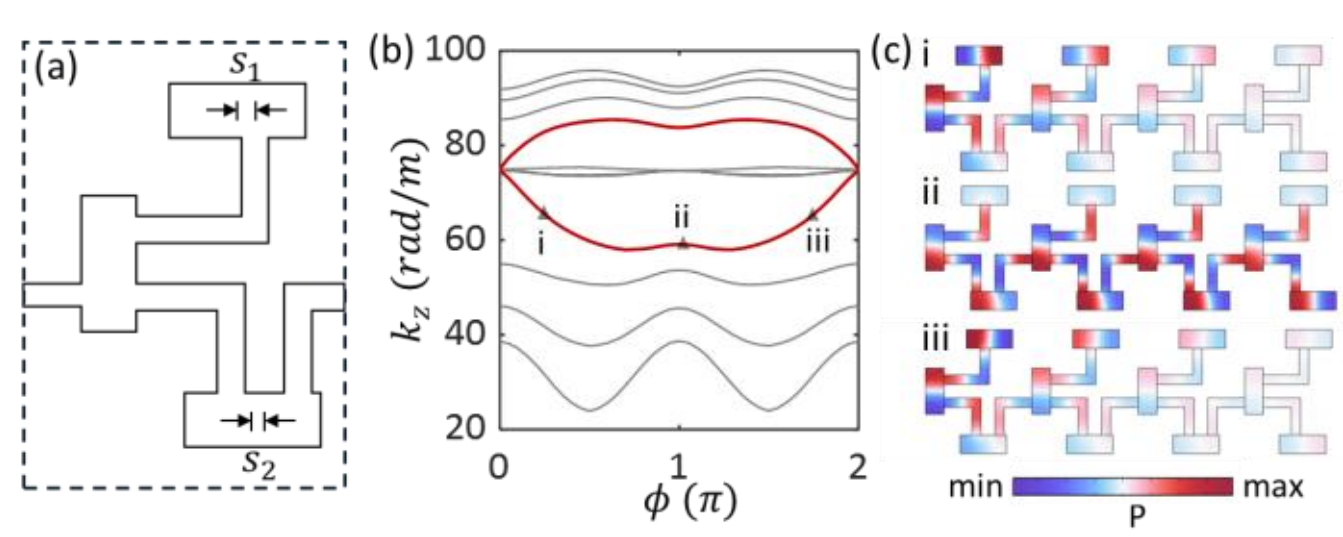

FIG. 3. (a) The schematic for the unit cell of the RTP acoustic waveguide array. $s_1$ is the displacement between the center of A waveguide and the connecting tube, while $s_2$ is the displacement between the center of C waveguide and the center of intra- and inter-cell connecting tubes. (b) The spectrum for the propagating constant $k_z$ as a function of evolution parameter $t$ at the frequency of 9500 Hz. (c) The simulated sound pressure distributions for the lower edge band with $t = \pi/4, \pi, 7\pi/4$.

Then, we consider a nontrivial evolution circle enclosing the dipole singularity with the form of

$$s_1 = 3.5\sin(t)\ [mm], s_2 = 2.5\cos(t)\ [mm], \qquad (5)$$

where $t$ is the evolution parameter. By treating the propagating distance $z$ as time, we plot the propagating constant $k_z$ as a function of $t$ in Fig. 3(b) for a finite waveguide array with four cells, which corresponds to the energy spectrum for the time-varying lattice in Fig. 2(d). The red curves at $t = 0$ represent a pair of edge states located at the left edge. The evolutionary feature for RTP can be revealed in a segment of the lower edge band from $t = \pi/4$ to $7\pi/4$ [labeled as points i and iii in Fig. 3(b)]. Fig. 3(c) illustrates the simulated sound pressure distributions for the points i ($t = \pi/4$), ii ($t = \pi$), and iii ($t = 7\pi/4$). At the initial point with $t = \pi/4$, the sound pressure energy locates at A and B waveguides at the left edge [see Fig. 3(c-i)]. The nearest ends of A and B waveguides have a $\pi$ phase difference, which can be defined as a $|{\downarrow}\rangle$ mode. With increase of $t$, the sound pressure penetrates into bulk [see Fig. 3(c-ii)] and finally returns to the original A and B waveguides at the left edge [see Fig. 3(c-iii)]. When $t = 7\pi/4$, the nearest ends of A and B waveguides have the same phase, which represents a $|{\uparrow}\rangle$ mode. Furthermore, the full-wave simulation for the whole RTP process [$t \in (0, 2\pi)$] with different input frequencies are presented in Section 6 of [52]. The above evolution process of sound pressure distribution provides a clear evidence of RTP. In addition, the sound pressure distributions for the upper edge band are presented in Section 7 of [52].

*Experimental results*. To realize the RTP process in the experiment, we modulate the displacement of A and C waveguides $s_{1,2}$ in the z direction, as shown in Fig. 4(a). The waveguide array is formed by a long evolutionary section $C_2$ with 1400 mm and short input/output sections $C_{1/3}$ with 70 mm. The relationship between the structure parameter $t$ and the propagating distance $z$ can be written as

$$\begin{aligned} C_1: t &= \frac{\pi}{4}, && z \in (0, 70)mm \\ C_2: t &= \frac{\pi}{4} + \frac{3\pi}{2}\left(\frac{z-70}{1400}\right), && z \in (70, 1470)mm \\ C_3: t &= \frac{7\pi}{4}. && z \in (1470, 1540)mm \end{aligned} \qquad (6)$$

Noting that this setup is above the adiabatic threshold (see details in Section 8 of [52]).

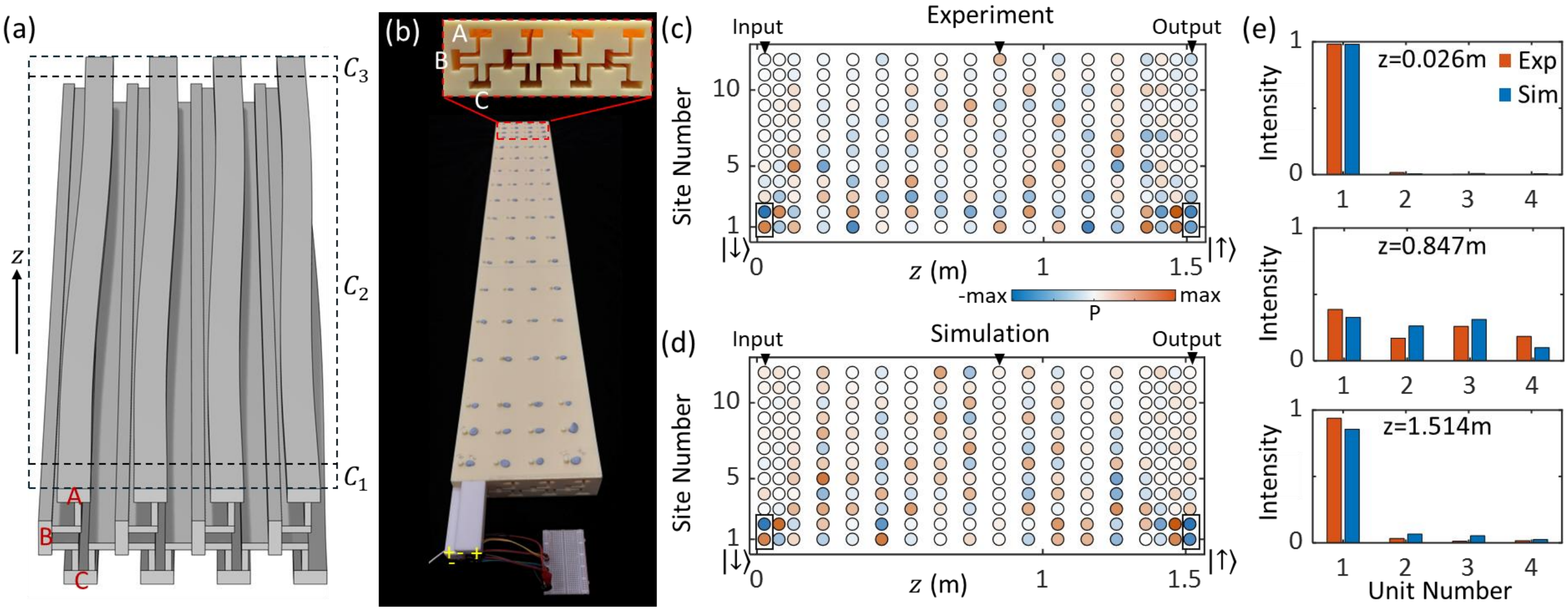

FIG. 4. (a) The configuration of the acoustic waveguide array. (b) The photo of the fabricated sample. (c, d) The measured and simulated sound pressure distribution in the waveguide array, respectively. (e) The measured (red bars) and simulated (blue bars) sound intensity distributions for the initial ($z = 0.026m$), intermediate ($z = 0.847m$), and final ($z = 1.514m$) states. The intensity in a unit is the sum of sound intensity in A, B, and C waveguides in this unit.

Fig. 4(b) shows the 3D-printed experimental sample, a resin block with the hollow waveguide array presented in Fig. 4(a). To excite the initial $|{\downarrow}\rangle$ state, we add two isolated waveguides touching A and B waveguides at the left edge of the input port, and place four out-of-phase loudspeakers near the ends of these two waveguides, whose phase configurations are labeled as "+" and "-" in Fig. 4(b). The exciting frequency is 9500 Hz. To measure the evolution of the sound pressure field along the propagation direction, an array of tiny holes is drilled into the top sides of each B waveguide and left sides of each A/C waveguide. The holes are sealed by 3D printed plugs to ensure air tightness and are only unplugged when being measured (see more details of the experiment in Section 9 of [52]). The measured sound pressure distribution is shown in Fig. 4(c). At the input port, the sound pressure localizes at A and B waveguides at the left edge with an antisymmetric distribution, corresponding to the $|\downarrow\rangle$ mode. With $z$ increases, the edge-localized $|\downarrow\rangle$ mode gradually evolves into delocalized bulk states and finally the edge-localized $|\uparrow\rangle$ mode at the left edge of the output port, symmetrically distributed at A and B waveguides. Together with the well-matched simulation results in Fig. 4(d), the experimental results provide strong evidence of RTP. In addition, we plot the sound intensity distributions of four cells for the initial ($z = 0.026m$), intermediate ($z = 0.847m$), and final ($z = 1.514m$) states, showing a good consistence between experiment and simulation. The experimental detection of $|\uparrow\rangle \rightarrow |\downarrow\rangle$ pumping process in the upper edge band is shown in Figure. S9 [52].

To sum up, we have constructed the dipole-mediated RTP in the 1D acoustic waveguide array and investigated the associated pumping dynamics. Measurements of the sound pressure distribution in the waveguides reveal the exotic evolution features of RTP: an edge-localized initial $|\downarrow\rangle$ state will delocalize into bulk in the sub-circle and return to the original edge as a $|\uparrow\rangle$ state in the other sub-circle. Our work paves the way for exploration of new physical phenomena and device applications based on higher-order singularity-mediated dynamical effects in classical systems.

S. Zhang acknowledges the support of New Cornerstone Science Foundation, the Research Grants Council of Hong Kong (STG3/E-704/23-N, AoE/P-502/20, 17309021). S. Liang acknowledges the support of the Research Grants Council of Hong Kong (UGC/FDS24/E04/21). J. Zhu acknowledges the support of National Natural Science Foundation of China (92263208), the Research Grants Council of Hong Kong (AoE/P-502/20).

Data availability—The data that support the findings of this article are openly available [53].

* These authors contributed equally to this work.
‡jiezhu@tongji.edu.cn
†shuzhang@hku.hk

[1] D. J. Thouless, Quantization of particle transport, Phys Rev B **27**, 6083 (1983).

[2] Q. Niu and D. J. Thouless, Quantised adiabatic charge transport in the presence of substrate disorder and many-body interaction, J Phys A Math Gen **17**, 2453 (1984).

[3] Q. Niu, Towards a quantum pump of electric charges, Phys Rev Lett **64**, 1812 (1990).

[4] B. L. Altshuler and L. I. Glazman, Pumping Electrons, Science **283**, 1864 (1999).

[5] F. Zhou, B. Spivak, and B. Altshuler, Mesoscopic Mechanism of Adiabatic Charge Transport, Phys Rev Lett **82**, 608 (1999).

[6] R. Citro and M. Aidelsburger, Thouless pumping and topology, Nature Reviews Physics **5**, 87 (2023).

[7] M. Lohse, C. Schweizer, O. Zilberberg, M. Aidelsburger, and I. Bloch, A Thouless quantum pump with ultracold bosonic atoms in an optical superlattice, Nat Phys **12**, 350 (2016).

[8] S. Nakajima, T. Tomita, S. Taie, T. Ichinose, H. Ozawa, L. Wang, M. Troyer, and Y. Takahashi, Topological Thouless pumping of ultracold fermions, Nat Phys **12**, 296 (2016).

[9] H.-I. Lu, M. Schemmer, L. M. Aycock, D. Genkina, S. Sugawa, and I. B. Spielman, Geometrical Pumping with a Bose-Einstein Condensate, Phys Rev Lett **116**, 200402 (2016).

[10] W. Ma, L. Zhou, Q. Zhang, M. Li, C. Cheng, J. Geng, X. Rong, F. Shi, J. Gong, and J. Du, Experimental Observation of a Generalized Thouless Pump with a

Single Spin, Phys Rev Lett **120**, 120501 (2018).

[11] Y. E. Kraus, Y. Lahini, Z. Ringel, M. Verbin, and O. Zilberberg, Topological States and Adiabatic Pumping in Quasicrystals, Phys Rev Lett **109**, 106402 (2012).

[12] M. Verbin, O. Zilberberg, Y. Lahini, Y. E. Kraus, and Y. Silberberg, Topological pumping over a photonic Fibonacci quasicrystal, Phys Rev B **91**, 64201 (2015).

[13] Y. Ke, X. Qin, F. Mei, H. Zhong, Y. S. Kivshar, and C. Lee, Topological phase transitions and Thouless pumping of light in photonic waveguide arrays, Laser Photon Rev **10**, 995 (2016).

[14] A. Cerjan, M. Wang, S. Huang, K. P. Chen, and M. C. Rechtsman, Thouless pumping in disordered photonic systems, Light Sci Appl **9**, 178 (2020).

[15] Z.-G. Chen, W. Tang, R.-Y. Zhang, Z. Chen, and G. Ma, Landau-Zener Transition in the Dynamic Transfer of Acoustic Topological States, Phys Rev Lett **126**, 54301 (2021).

[16] Z. Chen, Z. Chen, Z. Li, B. Liang, G. Ma, Y. Lu, and J. Cheng, Topological pumping in acoustic waveguide arrays with hopping modulation, New J Phys **24**, 013004 (2022).

[17] W. Song, O. You, J. Sun, S. Wu, C. Chen, C. Huang, K. Qiu, S. Zhu, S. Zhang, and T. Li, Fast topological pumps via quantum metric engineering on photonic chips, Sci Adv **10**, eadn5028 (2025).

[18] A. Cerjan, M. Wang, S. Huang, K. P. Chen, and M. C. Rechtsman, Thouless pumping in disordered photonic systems, Light Sci Appl **9**, 178 (2020).

[19] M. Lohse, C. Schweizer, H. M. Price, O. Zilberberg, and I. Bloch, Exploring 4D quantum Hall physics with a 2D topological charge pump, Nature **553**, 55 (2018).

[20] O. Zilberberg, S. Huang, J. Guglielmon, M. Wang, K. P. Chen, Y. E. Kraus, and M. C. Rechtsman, Photonic topological boundary pumping as a probe of 4D quantum Hall physics, Nature **553**, 59 (2018).

[21] Y. E. Kraus, Z. Ringel, and O. Zilberberg, Four-Dimensional Quantum Hall Effect in a Two-Dimensional Quasicrystal, Phys Rev Lett **111**, 226401 (2013).

[22] O. You, S. Liang, B. Xie, W. Gao, W. Ye, J. Zhu, and S. Zhang, Observation of Non-Abelian Thouless Pump, Phys Rev Lett **128**, 244302 (2022).

[23] Z.-G. Chen, R.-Y. Zhang, C. T. Chan, and G. Ma, Classical non-Abelian braiding of acoustic modes, Nat Phys **18**, 179 (2022).

[24] Y.-K. Sun, X.-L. Zhang, F. Yu, Z.-N. Tian, Q.-D. Chen, and H.-B. Sun, Non-Abelian Thouless pumping in photonic waveguides, Nat Phys **18**, 1080 (2022).

[25] X.-L. Zhang, F. Yu, Z.-G. Chen, Z.-N. Tian, Q.-D. Chen, H.-B. Sun, and G. Ma, Non-Abelian braiding on photonic chips, Nat Photonics **16**, 390 (2022).

[26] Y.-K. Sun, Z.-L. Shan, Z.-N. Tian, Q.-D. Chen, and X.-L. Zhang, Two-dimensional non-Abelian Thouless pump, Nat Commun **15**, 9311 (2024).

[27] W. Song, X. Liu, J. Sun, O. You, S. Wu, C. Chen, S. Zhu, T. Li, and S. Zhang, Shortcuts to adiabatic non-Abelian braiding on silicon photonic chips, Sci Adv **11**, eadt7224 (2025).

[28] M. Jürgensen, S. Mukherjee, and M. C. Rechtsman, Quantized nonlinear Thouless pumping, Nature **596**, 63 (2021).

[29] Q. Fu, P. Wang, Y. V Kartashov, V. V Konotop, and F. Ye, Nonlinear Thouless Pumping: Solitons and Transport Breakdown, Phys Rev Lett **128**, 154101 (2022).

[30] Q. Fu, P. Wang, Y. V Kartashov, V. V Konotop, and F. Ye, Two-Dimensional Nonlinear Thouless Pumping of Matter Waves, Phys Rev Lett **129**, 183901 (2022).

[31] M. Jürgensen and M. C. Rechtsman, Chern Number Governs Soliton Motion in Nonlinear Thouless Pumps, Phys Rev Lett **128**, 113901 (2022).

[32] N. Mostaan, F. Grusdt, and N. Goldman, Quantized topological pumping of solitons in nonlinear photonics and ultracold atomic mixtures, Nat Commun **13**, 5997 (2022).

[33] M. Jürgensen, S. Mukherjee, C. Jörg, and M. C. Rechtsman, Quantized fractional Thouless pumping of solitons, Nat Phys **19**, 420 (2023).

[34] S. Ravets, N. Pernet, N. Mostaan, N. Goldman, and J. Bloch, Thouless Pumping in a Driven-Dissipative Kerr Resonator Array, Phys Rev Lett **134**, 93801 (2025).

[35] A.-S. Walter, Z. Zhu, M. Gächter, J. Minguzzi, S. Roschinski, K. Sandholzer, K. Viebahn, and T. Esslinger, Quantization and its breakdown in a Hubbard–Thouless pump, Nat Phys **19**, 1471 (2023).

[36] A. Nelson, T. Neupert, T. Bzdušek, and A. Alexandradinata, Multicellularity of Delicate Topological Insulators, Phys Rev Lett **126**, 216404 (2021).

[37] A. Nelson, T. Neupert, A. Alexandradinata, and T. Bzdušek, Delicate topology protected by rotation

symmetry: Crystalline Hopf insulators and beyond, Phys Rev B **106**, 75124 (2022).

[38] P. Zhu, J. Noh, Y. Liu, and T. L. Hughes, Scattering theory of delicate topological insulators, Phys Rev B **107**, 195110 (2023).

[39] P. Zhu, A. Alexandradinata, and T. L. Hughes, ${\mathbb{Z}}_{2}$ spin Hopf insulator: Helical hinge states and returning Thouless pump, Phys Rev B **107**, 115159 (2023).

[40] Z.-Y. Zhuang, C. Zhang, X.-J. Wang, and Z. Yan, Berry-dipole semimetals, Phys Rev B **110**, L121122 (2024).

[41] A. Graf and F. Piéchon, Massless multifold Hopf semimetals, Phys Rev B **108**, 115105 (2023).

[42] D.-L. Deng, S.-T. Wang, C. Shen, and L.-M. Duan, Hopf insulators and their topologically protected surface states, Phys Rev B **88**, 201105 (2013).

[43] C. Liu, F. Vafa, and C. Xu, Symmetry-protected topological Hopf insulator and its generalizations, Phys Rev B **95**, 161116 (2017).

[44] T. Schuster, S. Gazit, J. E. Moore, and N. Y. Yao, Floquet Hopf Insulators, Phys Rev Lett **123**, 266803 (2019).

[45] T. Schuster, F. Flicker, M. Li, S. Kotochigova, J. E. Moore, J. Ye, and N. Y. Yao, Realizing Hopf Insulators in Dipolar Spin Systems, Phys Rev Lett **127**, 15301 (2021).

[46] Z. Wang, X.-T. Zeng, Y. Biao, Z. Yan, and R. Yu, Realization of a Hopf Insulator in Circuit Systems, Phys Rev Lett **130**, 57201 (2023).

[47] F. N. Ünal, A. Eckardt, and R.-J. Slager, Hopf characterization of two-dimensional Floquet topological insulators, Phys Rev Res **1**, 22003 (2019).

[48] F. N. Ünal, A. Bouhon, and R.-J. Slager, Topological Euler Class as a Dynamical Observable in Optical Lattices, Phys Rev Lett **125**, 53601 (2020).

[49] Q. Mo, R. Zheng, C. Lu, X. Huang, Z. Liu, and S. Zhang, Observation of Oriented Landau Levels and Helical Zero Modes in Berry Dipole Acoustic Crystals, Phys Rev Lett **134**, 116604 (2025).

[50] Q. Mo, S. Liang, C. Lu, J. Zhu, and S. Zhang, Tensor-Monopole-Induced Topological Boundary Effects in Four-Dimensional Acoustic Metamaterials, Phys Rev Lett **134**, 186601 (2025).

[51] A. Alexandradinata, Quantization of intraband and interband Berry phases in the shift current, Phys Rev B **110**, 75159 (2024).

[52] See Supplemental Material at [URL] for Hamiltonian, pseudospin reversal, simulation, and experiment.

[53] Q. Mo, S. Liang, X. Lan, J. Zhu, and S Zhang, 10.6084/m9.figshare.29965295.